
\input amstex
\documentstyle{amsppt}
\magnification=1200
\NoBlackBoxes
\NoRunningHeads
\define\slt{\operatorname{sl}(2,\Bbb C)}

\define\GL{\operatorname{GL}}
\define\G{\operatorname{G}}
\define\SUt{\operatorname{SU}(3)}
\define\Cvir{\operatorname{\Bbb Cvir}}
\define\Vrt{\operatorname{Vert}}
\document
\qquad\qquad\qquad\qquad
\qquad\qquad\qquad\qquad\qquad\qquad\qquad\qquad\qquad\qquad hep-th/9401047
\centerline{}
\topmatter
\title OCTONIONS AND BINOCULAR MOBILEVISION
\endtitle
\author
\centerline{}
\centerline{}
\centerline{}
\centerline{\qquad\qquad\qquad\qquad\qquad\qquad
{\rm In memory of beautiful days of my being at
Tartu.}}
\centerline{}
\centerline{}
\centerline{{\tt D.JURIEV}}
\centerline{}
\centerline{}
{\rm Mathematical Division,\linebreak Research Institute for System
Studies\linebreak
of Russian Academy of Sciences, ul.Avtozavodskaya 23,\linebreak
109280 Moscow, Rep. of Russia}\linebreak
\endauthor
\abstract
This paper is devoted to an interaction of two objects: the first of them is
octonions, the classical structure of pure mathematics, the second one is
Mobilevision, the recently developed technique of computer graphics. Namely,
it is shown that the binocular Mobilevision may be elaborated by use of the
octonionic colour space --- the seven dimensional extension of the classical
one, which includes a strange overcolour besides two triples of ordinary ones
(blue, green and red for left and right eyes).\newline
\centerline{}
\centerline{}
\centerline{}
\centerline{}
\centerline{}
Electronic mail: juriev\@systud.msk.su\newline
\centerline{}
\centerline{}
\endabstract
\endtopmatter
\newpage
\eightpoint
\head 1. INTERPRETATIONAL GEOMETRY, ANOMALOUS VIRTUAL REALITIES, QUANTUM
PROJECTIVE FIELD THEORY AND MOBILEVISION\endhead
\subhead 1.1. Interpretational geometry \endsubhead

A geometry being described below is related to a certain class of
{\it interactive information systems}.
Namely, let us call interactive information system {\it computerographical\/}
if the information stream from the computer is mounted as a stream of
geometric graphical data on a screen of the display; an interactive
computerographical information system is called {\it psychoinformation\/} one
if the information from observer to computer transmits unconsciousnessly
(such systems are widely applied f.e. in the medical diagnostics or
for the psychophysiological self-regulation, computer hypnosis and suggestion,
so below we shall be interested presumably in them).
We shall define conceptes of the interpretational figure and its symbolic
draws, which as it seems play a key role in the description of the
computer-geometric representation of mathematical data in interactive
information systems.

Mathematical data in interactive information system exist in the form of an
interrelation of {\it an interior geometric image\/} ({\it figure\/}) in the
subjective space of observer and {\it an exterior computerographical
representation\/} (cf.f.e.[1]).
The exterior computerographical representation includes {\it the visible
elements\/} (draws of figure) as well as of {\it the invisible ones\/}
(f.e. analytic expressions and algorythms of the constructing of such draws,
cf.[2]).
Process of the corresponding of a geometrical image (figure) in the interior
space of observer to a computerographical representation (visible and
invisible elements) will be called {\it translation}.
For example, a circle as a figure is a result of the translation of its draw on
a screen of the videocomputer (the visible object), constructed by an analytic
formula (the invisible object) accordingly to the fixed algorythm (also the
invisible object (cf.[3,4])).
It should be mentioned that the visible object may be nonidentical to the
figure, f.e. if a 3--dimensional body is defined by an axonometry, in three
projections, cross-sections or cuts, or in the window technique, which allows
to scale up a concrete detail of a draw (which is a rather pithy operation for
the visualization of fractals [5--8]), etc; in this case partial visible
elemnts may be regarded as modules, which translation is realized separately.
Continuing to use the terminology of computer science we shall call the
translation by {\it interpretation\/} if the translation of partial modules is
realized depending on the result of the translation of preceeding ones and by
{\it compilation\/} otherwise.
An example of the interpretation may be produced by the drawing of a fractal
which structure is defined by an observer on each step of the scaling up in
the window technique; the translation of visible elements in an intentional
anomalous virtual reality (see below) is also an interpretation.

\definition{Definition 1} A figure, which is obtained as a result of the
interpretation, will be called {\it interpretational figure\/}.\enddefinition

It should be mentioned that an interpretational figure may have no any
habitual formal definition; namely, only if the process of interpretation has
an equivalent compilation process then the definition of figure is reduced to
sum of definitions of its draws; nevertheless, in interactive information
systems not each interpretation process has an equivalent compilation one.
The draw of interpretational figure may be characterized only as "visual
perception technology" of figure but not as an "image", such draws will be
called {\it symbolic\/}.

The computer--geometric description of mathematical data in interactive
information systems is deeply related to the concept of anomalous virtual
reality.
It should be mentioned that there exist not less than two approaches to
foundations of geometry: in the first one the basic geometric objects are
figures defined by their draws, geometry describes realtions between them, in
the second one the basic geometric concept is a space (a medium, a field),
geometry describes various properties of a space and its states, which are
called the draws of figures.
For the purposes of the describing of geometry of interactive information
systems it is convenient to follow the second approach; the role of the medium
is played by an anomalous virtual reality, the draws of figures are its
certain states.

\subhead 1.2. Anomalous virtual realities \endsubhead

\definition{Definition 2 {\rm [9]}}

A. {\it Anomalous virtual reality\/} ({\it AVR\/}) {\it in a narrow sense\/}
is a certain system of rules of non--standard descriptive geometry adopted to
a realization on videocomputer (or multisensor system of "virtual reality").
{\it Anomalous virtual reality in a wide sense\/} contains also an image in
the cyberspace made accordingly to such system of rules.
We shall use this term in a narrow sense below.

B. {\it Naturalization\/} is the corresponding of an anomalous virtual reality
to an abstract geometry or a physical model.
We shall say that the anomalous virtual reality {\it naturalizes\/} the model
and such model {\it transcendizes\/} the naturalizing anomalous virtual
reality.

C. {\it Visualization\/} is the corresponding of certain images or visual
dynamics in the anomalous virtual reality to objects of the abstract geometry
or processes in the physical model.

D. An anomalous virtual reality, whose images depends on an observer, is
called {\it intentional anomalous virtual reality\/} ({\it IAVR\/}).
Generalized perspective laws in IAVR contain the equations of dynamics of
observed images besides standard (geometric) perspactive laws. A process of
observation in IAVR contains a physical process of observation and a virtual
process of intention, which directs an evolution of images accordingly to
dynamical laws of perspective.
\enddefinition

In the intentional anomalous virtual reality objects of observation present
themselves being connected with observer, who acting on them in some way,
determines, fixes their observed states, so an object is thought as a
potentiality of a state from the defined spectrum, but its realization depends
also on observer.
The symbolic draws of interpretational figures are presented by states of a
certain intentional anomalous virtual reality.

\subhead 1.3. Colours in anomalous virtual realities \endsubhead

It should be mentioned that the deep difference of descriptive geometry of
computerographical information systems from the classical one is the presense
of colours as important bearers of visual information.
The reduction to shape graphics, which is adopted in standard descriptive
geometry, is very inconvenient, since the use of colours is very familiar in
the scientific visualization [10--13].
The approach to the computerographical interactive information systems based
on the concept of anomalous virtual reality allows to consider an
investigation of structure of a colour space as a rather pithy problem of
descriptive geometry, because such space may be much larger than the usual one
and its structure may be rather complicated. Also it should be mentioned that
the using of other colour spaces allows to transmit diverse information in
different forms, so an investigation of the information transmission via
anomalous virtual realities, which character deeply depends on a structure of
colour space, become also an important mathematical problem (cf.[1]).

\definition{Definition 3}

A. A set of continuously distributed visual characteristics of image in an
anomalous virtual reality is called {\it anomalous colour space\/}. Elements
of an anomalous colour space, which have non--colour nature, are called {\it
overcolours\/}, and quantities, which transcendize them in the abstract model,
are called {\it "latent lights"\/}. {\it Colour--perspective system\/} is a
fixed set of generalized perspective laws in fixed anomalous colour space.

B. The transmission of information via anomalous virtual reality by "latent
lights" is called {\it AVR--photodosy\/}.
\enddefinition

\subhead 1.4. Quantum projective field theory \endsubhead

Mobilevision is an intentional anomalous virtual reality naturalizing {\it the
quantum projective field theory\/} ([14,15]). The process of naturalization is
described in [10,15]. Its key points will be presented below, here our
attention is concentrated on the basic concepts of the quantum projective
field theory, which naturalization Mobilevision is.

\definition{Definition 4A {\rm [14,15]}} {\it QFT--operator algebra\/} ({\it
operator algebra of the quantum field theory, vertex operator algebra, vertex
algebra\/}) is the pair $(H,t^k_{ij}(\vec x))$: $H$ is a linear space,
$t^k_{ij}(\vec x)$ is $H$--valued tensor field such that $t^l_{im}(\vec
x)t^m_{jk}(\vec y)\!=\!t^m_{ij}(\vec x\!-\!\vec y)t^l_{mk}(\vec y)$.
\enddefinition

Let us intruduce the operators $l_{\vec x}(e_i)e_j=t^k_{ij}(\vec x)e_k$, then
the following relations will hold: $l_{\vec x}(e_i)l_{\vec
y}(e_j)=t^k_{ij}(\vec x-\vec y)l_{\vec y}(e_k)$ ({\it operator product
expansion\/}) and $l_{\vec x}(e_i)l_{\vec y}(e_j)=l_{\vec y}(l_{\vec x-\vec
y}(e_i)e_j)$ ({\it duality\/}).

\definition{Definition 4B {\rm [14,15]}}
QFT--operator algebra $(H,t^k_{ij}(u); u\in\Bbb C)$ is called {\it
QPFT--operator
algebra\/} ({\it operator algebra of the quantum projective field theory\/})
iff (1) $H$ is the sum of Verma modules $V_{\alpha}$ over $\slt$ with the
highest vectors $v_{\alpha}$ and the highest weights $h_{\alpha}$, (2)
$l_u(v_{\alpha})$ is a primary field of spin $h_{\alpha}$, i.e.
$[L_k,l_u(v_\alpha)]=(-u)^k(u\partial_u+(k+1)h_{\alpha})l_u(v_{\alpha})$,
where $L_k$ are the $\slt$ generators ($[L_i,L_j]=(i-j)L_{i+j}$,
$i,j=-1,0,1$),
(3) the descendant generation rule holds: $L_{-1}l_u(f)=l_u(L_{-1}f)$.
QFT--operator algebra $(H,t^k_{ij}(u); u\in\Bbb C)$ is called {\it derived
QPFT--opera-\linebreak tor algebra\/} iff conditions (1) and (2) as well as
derived rule
of descendants generation ($[L_{-1},l_u(f)]=l_u(L_{-1}f)$) hold.
\enddefinition

As it was shown in the paper [14] the categories of QPFT--operator algebras
and derived QPFT--operator algebras are equivalent.
The explicit construction of equivalence was described.
Therefore, QPFT--operator algebras and derived QPFT--operator algebras may be
considered as different recordings of the same object, and one may use the
most convenient one in each concrete case.

It should be mentioned also that in arbitrary QFT--operator algebra one can
define an operation depending on the parameter: $m_u(e_i,e_j)=t^k_{ij}(u)e_k$.
For this operation the following identity holds:
$m_u(\cdot,m_v(\cdot,\cdot))=m_v(m_{u-v}(\cdot,\cdot),\cdot)$.
The operators $l_u(f)$ are the operators of the left multiplication in the
obtained algebra.

\definition{Definition 4C {\rm [15]}}
QPFT--operator algebra (derived QPFT--operator algebra) $(H,t^k_{ij}(u))$ is
called {\it projective $G$--hypermultiplet\/}, iff the group $G$ acts in it by
automorphisms, otherwords, the space $H$ possesses a structure of the
representation of the group $G$, the representation operators commute with the
action of $\slt$ and $l_u(T(g)f)=T(g)l_u(f)T(g^{-1})$.
\enddefinition

The linear spaces of the highest vectors of the fixed weight form
subrepresentations of $G$, which are called {\it multiplets\/} of projective
$G$-hypermultiplet.

\subhead 1.5. Mobilevision \endsubhead

As it was mentioned above {\it Mobilevision\/} is a certain anomalous virtual
reality, which naturalizes the quantum projective field theory.
Possibly, Mobilevision is not its unique naturalization.
Here we describe the key moments of the process of naturalization of the
quantum projective field theory which is resulted in Mobilevision.

Unless the abstract model (quantum projective field theory) has a quantum
character the images in its naturalization, the intentional anomalous virtual
reality of Mobilevision, are classical.
The transition from the quantum field model to classical one is done by
standard rules [16], namely, the classical field with Taylor coefficients
$|a_k|^2$ is corresponded to the element $\sum a_k L_{-1}^k v_{\alpha}$ of the
QPFT--operator algebra.

Under the naturalization three classical fields are identified with fields of
three basic colours (red, green and blue -- see f.e. [17]), other fields with
fields of overcolours; there are pictured only the colour characteristics for
the fixed moment of time on the screen of the videocomputer as well as the
perception of the overcolours by an observer is determined by the intentional
character of the anomalous virtual reality of Mobilevision.
Namely, during the process of the evolution of the image, produced by the
observation, the vacillations of the colour fields take place in accordance to
the dynamical perspective laws of the intentioanl anomalous virtual reality of
Mobilevision (Euler formulas [9] or Euler--Arnold equations [15]).
These vacillations depend on the character of an observation (f.e. the eye
movement or another dynamical parameters); the vacillating image depends on
the distribution of the overcolours, that allows to interpret the overcolours
as certain vacillations of the ordinary colours.
So the overcolours in the intentional anomalous virtual reality of
Mobilevision are vacillations of the fixed type and structure of ordinary
colours with the defined dependence on the parameters of the observation
process.
The transcending "latent lights" are the quantized fields of the basic model
of the quantum projective field theory.

The presence of the $\SUt$--symmetry of classical colour space (see 3.1)
allows to suppose that the QPFT--operator algebra of the initial model is the
projective $\SUt$--hypermultiplet.

Now we are interested in the investigation of a behaviour of overcolours under
scaling transformations, the extraction of natural classes of overcolours with
respect to these transformations and describe some properties of these
classes.

\proclaim{Proposition 1} If the initial quantized field of "latent light" in
the abstract model of the quantum projective field theory has the spin $h$,
then the lightening of the corresponding overcolour in the intentional
anomalous virtual reality of Mobilevision increase in $s^{2h}$ times under the
scaling up in $s$ times.
\endproclaim

This statement follows from the description of the naturalization process
given above.

Therefore, there are the currents, quantized fields of spin 1, corresponded to
the ordinary colours.
Let us mention that for the fields of "latent lights" of negative spin
({\it ghosts\/}) the anomalous increasing of lightening holds under
the scaling down i.e. the moving of object away from observer.
So the virtual diffusion of ghosts in cyberspace holds with an intensification
accordingly to a principle of a snow--slip.
There are the overcolours, which lightening desease under the scaling down
(i.e. the moving of object away from observer) quicker then for ordinary
colours, corresponded to "latent lights" of spin greater than 1.

\head 2. QUANTUM CONFORMAL AND $q_R$--CONFORMAL FIELD THEORIES, AN INFINITE
DIMENSIONAL QUANTUM GROUP AND QUANTUM--FIELD ANALOGS OF EULER--ARNOLD
EQUATIONS
\endhead

\subhead 2.1. Quantum conformal field theory
\endsubhead

\definition{Definition 7A {\rm [18]}}

A. The highest vector $T$ of the weight 2 in the QPFT--opeartor algebra will
be called {\it the conformal stress--energy tensor\/} if $T(u):=l_u(T)=\sum L_k
(-u)^{k-2}$, where the operators $L_k$ form the Virasoro algebra:
$[L_i,L_j]=(i-j)L_{i+j}+\frac{i^3-i}{12} c\cdot I$.

B. The set of the highest vectors $J^\alpha$ of the weight 1 in the
QPFT--operator algebra will be called the set of {\it the affine currents\/}
if $J^\alpha(u):=l_u(J^\alpha)=\sum J^\alpha_k(-u)^{k-1}$, where the
operators $J^\alpha_k$ form {\it the affine Lie algebra\/}:
$[J^\alpha_i,J^\beta_j]=c^{\alpha\beta}_\gamma
J^\gamma_{i+j}+k^{\alpha\beta}\cdot i\delta(i+j)\cdot I$.
\enddefinition

In view of the results of [19,20] the quantum projective field theories with
conformal stress--energy tensor are just conformal field theories in sense of
[21,22].

If there is defined a set of the affine currents in the QPFT--operator algebra
then one can construct the conformal stress--energy tensor by use of {\it
Sugawara construction\/} [23] or more generally by use of {\it the Virasoro
master equations\/} [24--26].

Below we shall be interested in the special deformations of the quantum
conformal field theories in class of the quantum projective ones
(cf.[27--29]), which will be called quantum $q_R$--conformal field theories.

\subhead 2.2. Lobachevskii algebra, the quantization of the Lobachevskii plane
\endsubhead

In the Poincare realization of the Lobachevskii plane (the realization in the
unit disk) the Lobachevskii metric may be written as follows:
$$w=q_R^{-1}\,dzd\bar{z}/(1-|z|^2)^2.$$

We should construct the $C^*$--algebra, which may be considered as a
quantization of such metric, namely, let us consider two variables $t$ and
$t^*$, which obey the following commutation relations: $[tt^*,t^*t]=0$,
$[t,t^*]=q_R(1-tt^*)(1-t^*t)$. One may realize such variables by bounded
operators in the Verma module over $\slt$ of the weight
$h=\frac{q_R^{-1}+1}2$.
If such Verma module is realized in polynomials of one complex variable $z$
and the action of $\slt$ has the form $L_{-1}=z$, $L_0=z\partial_z+h$,
$L_1=z(\partial_z)^2+2h\partial_z$, then the variables $t$ and $t^*$ are
represented by tensor operators $\partial_z$ and $z/(z\partial_z+2h)$.
These operators are bounded if the Verma module is unitarizable
($h>\frac12$, $q_R>0$) and therefore one can construct a Banach algebra
generated by them and obeying the prescribed commutation relations.
The structure of $C^*$--algebra is rather obvious: an involution $*$ is
defined on generators in a natural way, because the corresponding tensor
operators are conjugate to each other.
\pagebreak

\subhead 2.3. Quantum $q_R$--conformal field theory \endsubhead

\definition{Definition 7B {\rm [18]}}

A. The highest vector $T$ of the weight 2 in the QPFT--operator algebra will
be called {\it the $q_R$--conformal stress--energy tensor\/} if
$T(u):=l_u(T)=\sum L_k (-u)^{k-2}$, where the operators $L_k$ form the
$q_R$--Virasoro algebra:
$$\align
[L_i,L_j]=&(i-j)L_{i+j}\qquad\qquad (i,j\ge -1;\quad i,j\le 1),\\
[L_2,L_{-2}]=&\varphi(L_0+1)-\varphi(L_0-1),\\
\varphi(t)=&\frac{t(t+1)(t+3h-1)^2}{(t+2h)(t+2h-1)}.
\endalign
$$

B. The set of the highest vectors $J^\alpha$ of the weight 1 in the
QPFT--operator algebra will be called the set of {\it the $q_R$--affine
currents\/} if $J^\alpha(u):=l_u(J^\alpha)=\sum J^\alpha_k(-u)^{k-1}$, where
the operators $J^\alpha_k$ form {\it the $q_R$--affine Lie algebra\/}:
$$\align J^{\alpha}_k=&J^{\alpha}T^{-k}f_k(t),\,
[J^{\alpha},J^{\beta}]=c^{\alpha\beta}_{\gamma},\\
Tf(t)=&f(t+1)T,\,
[T,J^{\alpha}]=[f(t),J^{\alpha}]=0,\\
f_k(t)=&t\ldots(t-k),\text{ if } k\ge 0,\text{ and }
((t+2h)\ldots (t+2h-k+1))^{-1},\text{ if } k\le 0,\\
h=&(q^{-1}_R+1)/2.\endalign
$$
\enddefinition

It should be mentioned that $q_R$--affine currents and $q_R$--conformal
stress--energy tensor are just the $\slt$--primary fields in the Verma module
$V_h$ ($h=\frac{q^{-1}_R+1}2$) over $\slt$ of spin 1 and 2, respectively.
If such module is realized as before then
$$\align
J_k=&\partial_z^k,\, J_{-k}=z^k/(\xi+2h)\ldots(\xi+2h+k-1);\\
L_2=&(\xi+3h)\partial^2_z,\, L_1=(\xi+2h)\partial_z,\, L_0=\xi+h,\,
L_{-1}=z,\\
L_{-2}=&z^2\frac{\xi+3h}{(\xi+2h)(\xi+2h+1)},\, \xi=z\partial_z;\endalign$$
the $q_R$--affine algebra may be realized in terms of Lobachevskii
$C^*$--algebra:
$$J^\alpha_k=J^\alpha t^k,\text{ if } k\ge 0,\text{ and }
J^\alpha(t^*)^{-k},\text{ if } k\le 0,\quad
[J^\alpha,J^\beta]=c^{\alpha,\beta}_{\gamma}J^{\gamma}.$$
The QPFT--operator algebras generated by $q_R$--conformal currents are called
{\it canonical projective $G$--hypermultiplets\/} [15].

\subhead 2.4. An infinite dimensional quantum group \endsubhead

It should be mentioned that the primary fields $V_k(u)$ of nonnegative integer
spins $k$ in the Verma module $V_h$ [30,31], which form a closed
QPFT--operator algebra (as well as in the case of extended conformal field
theories [32--35], the components of the $\slt$--primary fields form in some
sense an analogue of conformal $W$--algebra [36--39]), are not local.
Nevertheless, their commutation relations may be described as follows
$$V_\alpha(u)V_\beta(v)=
S^{\gamma\delta}_{\alpha\beta}(u-v)V_\gamma(v)V_\delta(u).$$
That means that these primary fields form a Zamolodchikov algebra [40--42].
$S$--matrix $S^{\gamma\delta}_{\alpha\beta}(u;q_R)$ of this Zamolodchikov
algebra defines an infinite dimensional quantum group, which is a certain
deformation of $\GL(\infty)$, in a standard way [43,44].

\subhead 2.5. Quantum--field Euler--Arnold top and Virasoro master equation
\endsubhead

Let $H$ be an arbitrary direct sum of Verma modules over $\slt$ and $P$ be a
trivial fiber bundle over $\Bbb C$ with fibers isomorphic to $H$.
It should be mentioned that $P$ is naturally trivialized and possesses a
structure of $\slt$--homogeneous bundle.
A $\slt$--invariant Finsler connection $A(u,\partial_t u)$ in $P$ is called an
angular field [9].
Angular field $A(u,\partial_t u)$ may be expanded by $(\partial_t u)^k$, the
coefficients of such expansion are just $\slt$--primary fields [9].
The equation
$$\partial_t\Phi_t=A(u,\partial_t u)\Phi_t,$$
where $\Phi_t$ belongs to $H$ and $u=u(t)$ is the function of scanning, is a
quantum--field analog of the Euler formulas [9].
Such analog describes an evolution of Mobilevision image under the
observation.
One may also consider an affine version of the Euler formulas, which may be
written as follows
$$\partial_t\Phi_t=A(u,\partial_t u)(\Phi_t-\Phi_0).$$
Regarding canonical projective $G$--hypermultiplet we may construct a quantum
field analog of the Euler--Arnold equation [15]
$$\partial_t A=\{H,A_t\},$$
where an angular field $A(u,\partial_t u)$ is considered as an element of the
canonical projective $G$--hypermultiplet being expanded by $\slt$--primary
fields of this hypermultiplet, $H$ is the quadratic element of $S(\frak g)$,
$\{\cdot,\cdot\}$ are canonical Poisson brackets in $S(\frak g)$.
It is quite natural to demand $H$ be a solution of the Virasoro master
equation.
If we consider a projective $G$--hypermultiplet, which is a semi--direct
product of the canonical one and a trivial one (i.e. with $l_u(f)=0$), then it
will be possible to combine Euler--Arnold equations with Euler formulas to
receive the complete dynamical perspective laws of the Mobilevision.
Such construction will be used in the naxt paragraph for the description of
the octonionic colour space of binocular Mobilevision.

\head 3. OCTONIONIC COLOUR SPACE AND BINOCULAR MOBILEVISION \endhead

\subhead 3.1. Quaternionic description of ordinary colour space \endsubhead

It should be mentioned that ordinary colour space may be described by use of
imaginary quaternions in the following way: let us consider an arbitrary
imaginary complex quaternion $x=ri+bj+gk$, $i, j, k$ are imaginary roots and
$r, g, b$ are complex numbers.
One may correspond to such quaternion an element of the colour space, which in
RGB--coordinates [17] has components $R=|r|^2$, $G=|g|^2$, $B=|b|^2$.
The lightening $L$ has the quadratic form in the quaternionic space, namely,
$L=\frac12(|r|^2+|b|^2+|g|^2)$.
The group $\SUt$ is a group of its invariance.

\subhead 3.2. Octonionic colour space and binocular Mobilevision \endsubhead

Let us construct an octonionic colour space to describe the binocular
Mobilevision.
This space will be a semi--direct product of a canonical projective
$\G_2$--hypermultiplet on the trivial one, which is a direct sum of seven
copies of the suitable Verma module over $\slt$.
The group $\G_2$ acts in this seven dimensional space as it acts on imaginary
octonions [45] (see also [46,47]).
There is uniquely defined up to a multiple and modulo the Poissonic center an
$\SUt$--invariant quadratic element in $S^2(\frak g_2)$.
Moreover, it obeys the Virasoro master equation (as a solution of
$\G_2/\SUt$--coset model).
So we can construct the Euler--arnold equations in the canonical projective
$\G_2$--hypermultiplet.
To receive the binocular version of the affine Euler formulas one should use
the decomposition of $S^2(\frak g_2)$ on the $\SUt$--chiral components (left
and right); the angular fields from the chiral components will depend on
chiral parameters $u_l$, $\partial_t u_l$ and $u_r$, $\partial_t u_r$,
attributed to the left and right eyes, respectively.
Six copies of Verma modules over $\slt$, mentioned above, form a pair of
projective $\SUt$--hypermultiplets, which correspond to ordinary colours for
left and right eyes; one copy form also a projective $\SUt$--hypermultiplet,
its overcolour will be called {\it a strange overcolour}.
So the constructed seven dimensional octonionic colour space includes a pair
of ordinary three dimensional colour spaces (for left and right eyes,
respectively) and one strange overcolour.

\Refs

\roster
\item" [1]" Saati T.L., Speculating on the future of Mathematics. Appl. Math.
Lett. 1 (1988), 79-82.
\item" [2]" Turbo Vision for Pascal. IVC--Soft, Moscow, 1992 [in Russian].
\item" [3]" Gelfand I.M., Glagoleva E.G., Schnol E.E., Functions and graphics.
Nauka, Moscow, 1973 [in Russian].
\item" [4]" Gelfand I.M., Glagoleva E.G., Kirillov A.A., Coordinate method.
Nauka, Moscow, 1973 [in Russian].
\item" [5]" Mandelbrot B., Fractals. Form, chance and dimension. Freeman, San
Francisco, 1977.
\item" [6]" Mandelbrot B., The fractal geometry of nature. Freeman, San
Francisco, 1982.
\item" [7]" Fractal geometry and computer graphics. Ed. J.L.Encarna\~cao.
Springer, 1992.
\item" [8]" Peiten H.O., J\"urgens H., Saupe D., Fractals for the classroom.
Springer, 1992.
\item" [9]" Juriev D.V., Quantum projective field theory: quantum analogs of
the Euler formulas. Teor. Matem. Fiz. 92:1 (1992), 172-176 [in Russian].
\item"[10]" Visualization in scientific computing. Eds. G.M.Nielson and
B.Shriver, IEEE Comput. Soc. Press, Los Alamitos, 1989.
\item"[11]" Proc. Visualization 90. IEEE Comput. Soc. Press, Los Alamitos,
1990.
\item"[12]" Proc. Visualization 91. IEEE Comput. Soc. Press, Los Alamitos,
1991.
\item"[13]" Encarna\~cao J.L., Astheiner P., Felger W., Fruhauf M., G\"obel
M., Karlsson K., Graphics modelling as a basic tool for scientific
visualization. In "Modelling in computer graphics". Ed. T.L.Kunii, Springer,
1991.
\item"[14]" Bychkov S.A., Juriev D.V., Three algebraic structures of the
quantum projective field theory. Teor. Matem. Fiz. 97:3 (1993) [in Russian].
\item"[15]" Juriev D.V., Quantum projective field theory: quantum--field
analogs of the Euler--arnold equations in projective $G$--hypermultiplets.
Teor. Matem. Fiz. (1994) [in Russian].
\item"[16]" Bogolubov N.N., Shirkov D.V., Introduction to the theory of
quantized fields. Nauka, Moscow, 1976 [in Russian].
\item"[17]" Greenberg D., Joblove G.H., Color spaces for computer graphics.
Comput. Graph.\linebreak  SIGGRAPH--ACM, 12:3 (1978).
\item"[18]" Belavin A.A., Polyakov A.M., Zamolodchikov A.B., Infinite
conformal symmetry in two--dimensional quantum field theory. Nucl. Phys. B241
(1984), 333-380.
\item"[19]" Mack G., Introduction to conformal invariant quantum field theory
in two and more dimensions. In "Nonperturabative quantum field theory". Eds.
G.t'Hooft et al., Plenum, New York, 1988.
\item"[20]" Hadjivanov L.K., Existence of primary fields as a generalization
of the Luscher--Mack theorem. J. Math. Phys. 34 (1993), 441-453.
\item"[21]" Juriev D.V., Algebra $\Vrt(\Cvir;c)$ of the vertex operators for
the Virasoro algebra. Algebra i Anal. 3:3 (1991), 197-205 [in Russian].
\item"[22]" Juriev D.V., Quantum conformal field theory as infinite
dimensional noncommutative geometry. Uspekhi Matem. Nauk 46:4 (1991), 115-138
[in Russian].
\item"[23]" Goddard P., Olive D., Kac--Moody and Virasoro algebras in relation
to quantum physics. Intern. J. Mod. Phys. A1 (1986), 303-414.
\item"[24]" Halpern M.B., Kiritsis E., General Virasoro construction on affine
$\frak g$. Mod. Phys. Lett. A4 (1989), 1373-1380, 1797.
\item"[25]" Morozov A.Yu., Perelomov A.M., Rosly A.A., Shifman M.A., Turbiner
A.V., Quasi--exactly--solvable quantal problems: one--dimensional analogue of
rational conformal field theories. Intern. J. Mod. Phys. A5 (1990), 803-832.
\item"[26]" Halpern M.B., Recent developments in the Virasoro Master Equation.
Berkeley preprint, UCB-PTH-91/43 (1991).
\item"[27]" Zamolodchikov A.B., Higher integrals of motion in two--dimensional
field theories with broken conformal symmetry. Pis'ma ZhETP 46:4 (1987),
129-132 [in Russian].
\item"[28]" Zamolodchikov A.B., Integrable field theory from conformal field
theory. Proc. Taniguchi Symposium, Kyoto, 1988.
\item"[29]" Gerasimov A., Lebedev D., Morozov A., On possible implications of
2--dimensional integrable systems for string theory. Preprint ITEP 4-90
(1990).
\item"[30]" Juriev D., The explicit form of the vertex operator fields in
two--dimensional $\slt$--invariant field theory. Lett. Math. Phys. 22 (1991),
141-144.
\item"[31]" Juriev D.V., Classification of the vertex operators in
two--dimensional $\slt$--invariant field theory. Teor. Matem. Fiz. 86:3
(1991), 338-343 [in Russian].
\item"[32]" Fateev V.A., Lukyanov S.L., The models of two--dimensional
conformal quantum field theory with $\Bbb Z_n$--symmetry. Intern. J. Mod.
Phys. A3 (1988), 507-520.
\item"[33]" Bilal A., Gervais J.-L., Conformal theories with non--linearly
extended Virasoro symmetries and Lie algebra classification. In "Infinite
dimensional Lie algebras and Lie groups". Ed. V.Kac, Singapore, World
Scientific, Adv. Ser. Math. Phys. 7 (1989), 483-526.
\item"[34]" Bowcock P., Goddard P., Coset constructions and extended conformal
algebras. Nucl. Phys. B305 (1988), 685.
\item"[35]" Bilal A., Gervais J.-L., Systematic construction of conformal
theories with higher--spin Virasoro symmetries. Nucl. Phys. B318 (1989),
597-630.
\item"[36]" Bais F.A., Bouwknegt P., Surridge M., Schoutens K., Extensions of
the Virasoro algebra constructed from Kac--Moody algebras using higher order
Casimir invariants. Nucl. Phys. B304 (1988), 348-370.
\item"[37]" Lukyanov S.L., Quantization of the Gelfand-Dikii algebra. Funkt.
Anal. i ego Prilozh. 22:4 (1988), 1-10 [in Russian].
\item"[38]" Bakas I., Higher spin fields and the Gelfand-Dikii algebra.
Commun. Math. Phys. 123 (1989), 627-640.
\item"[39]" Nahm W., Conformal quantum field theory in two dimensions. World
Scientific, Singapore, 1992.
\item"[40]" Zamolodchikov A.B., Strict two--particle $S$--matrix of quantum
solitons in Sine--Gordon model. Pis'ma ZhETP 25:10 (1977), 499-502 [in
Russian].
\item"[41]" Sklyanin E.K., Faddeev L.D., Quantum mechanical approach to
integrable models of quantum field theory. Dokl. Akad. Nauk SSSR 243:6 (1978),
1430-1433 [in Russian].
\item"[42]" Takhtajan L.A., Faddeev L.D., Quantum inverse scattering method
and Heisenberg XYZ--model. Uspekhi Matem. Nauk 34:5 (1979), 13-63 [in
Russian].
\item"[43]" Reshetikhin N.Yu., Takhtajan L.A., Faddeev L.D., Quantization of
Lie groups and Lie algebras. Algebra i Anal. 1:1 (1989), 178-206 [in Russian].
\item"[44]" Manin Yu.I., Quantum groups and non--commutative geometry.
Preprint CRM-1561, Montr\'eal, 1988.
\item"[45]" Freudenthal H., Oktaven, Ausnahmengruppen und Oktavengeometrie.
Geom. Dedicata 19 (1985), 7-63.
\item"[46]" Sorgsepp L., L\^ohmus J., About nonassociativity in physics and
Cayley--Graves octonions. Hadronic J. 2 (1979), 1388-1459.
\item"[47]" L\^ohmus J., Sorgsepp L., Nonassociative algebras in physics.
Preprints Estonian Inst. Phys. Tartu; F-24,25, 1985 [in Russian].
\endroster
\endRefs
\enddocument